# Off-the-shelf deep learning is not enough: parsimony, Bayes and causality

Rama K. Vasudevan*[1], Maxim Ziatdinov[2], Lukas Vlcek[3,†] and Sergei V. Kalinin[1,**]

[1]Center for Nanophase Materials Sciences, [2]Computational Sciences and Engineering Division, and [3]Materials Science and Technology Division, Oak Ridge National Laboratory, Oak Ridge TN 37831, USA †Current Affiliation: Bayer, St. Louis, MO 63141

**Abstract**

Deep neural networks ('deep learning') have emerged as a technology of choice to tackle problems in natural language processing, computer vision, speech recognition and gameplay, and in just a few years has led to super-human level performance and ushered in a new wave of 'AI'. Buoyed by these successes, researchers in the physical sciences have made steady progress in incorporating deep learning into their respective domains. However, such adoption brings substantial challenges that need to be recognized and confronted. Here, we discuss both opportunities and roadblocks to implementation of deep learning within materials science, focusing on the relationship between correlative nature of machine learning and causal hypothesis driven nature of physical sciences. We argue that deep learning and AI are now well positioned to revolutionize fields where causal links are known, as is the case for applications in theory. When confounding factors are frozen or change only weakly, this leaves open the pathway for effective deep learning solutions in experimental domains. Similarly, these methods offer a pathway towards understanding the physics of real-world systems, either via deriving reduced representations, deducing algorithmic complexity, or recovering generative physical models. However, extending deep learning and 'AI' for models with unclear causal relationship can produce misleading and potentially incorrect results. Here, we argue the broad adoption of Bayesian methods incorporating prior knowledge, development of DL solutions with incorporated physical constraints, and ultimately adoption of causal models, offers a path forward for fundamental and applied research. Most notably, while these advances can change the way science is carried out in ways we cannot imagine, machine learning is not going to substitute science any time soon.

*vasudevanrk@ornl.gov, **sergei2@ornl.gov

*Introduction*

The spectacular growth of deep learning (DL) in the last decade has fueled the rise of a new wave of data science and artificial intelligence ('AI') that has already had global impact across society. The spectacular successes of deep learning in some traditionally very difficult tasks in computer vision, natural language processing, machine translation, speech recognition and gameplay has piqued interest across all scientific communities.[1] Here, deep learning refers to an approach that utilizes artificial neural networks (which have been available for decades[2]) that are comprised of numerous layers of stacked artificial neurons, and with oftentimes millions of trainable parameters, usually to approximate some highly complex nonlinear function. The networks are usually trained using a backpropagation algorithm and stochastic gradient descent to adjust the weights of the network to minimize some objective function, and nowadays expressly run on graphical or tensor processing units optimized for such calculations.

Depending on the specific architectures involved, deep neural networks (DNN) can be used in tasks including classification, regression (for instance, material property predictions based on a material's structure), as well as for unearthing correlations and compressing data in large datasets. A simple example of a DNN used in a materials science setting is shown in Figure 1: in this case this network has been 'trained' to automatically identify atoms from noisy electron microscopy images. The network was trained by ingesting large volumes of simulated electron microscopy images where the atomic positions are known and therefore used as the 'labeled' data. In this process the network's parameters are continually updated to minimize the discrepancy between the predictions of the network and the ground truth (the positions of the atoms). The network can then be fed a new image that was not part of the original training set to give the output of the atomic coordinates present, thereby operating as an automatic atom finder. In addition to simple image segmentation tasks, DNNs have also seen success in the trickier task of "generative modeling," which refers to the ability to generate new datapoints (samples) that are not in the original dataset.[3]

The key distinction between traditional machine learning (ML) and modern deep learning is that deep neural networks learn representations ('features') of the data as part of the training process, as opposed to being hand-crafted by domain experts, which was the prevailing method prior to the DL revolution. However, this also presents a problem: are the representations learned by the existing DL methods useful for aiding in understanding of physics and materials science? Even from a computer science perspective, DL, for all its successes, is surprisingly fragile and highly susceptible to adversarial attacks,[4] in which input data are slightly perturbed in subtle ways that slowly guide the network to mis-classify the data with near 100% certainty.[5] A recent example shows that a DL-trained classifier of objects can mis-classify simple objects merely if they are displayed in specific unseen poses.[6] How can we then 'trust' the predictions of DL-based models, when they appear highly fragile and vulnerable? Perhaps as a less exotic example, how do we know which network architecture will give a correct, quantitative answer for a specific problem, and how can we quantify uncertainties and systematic and random errors in such an answer?

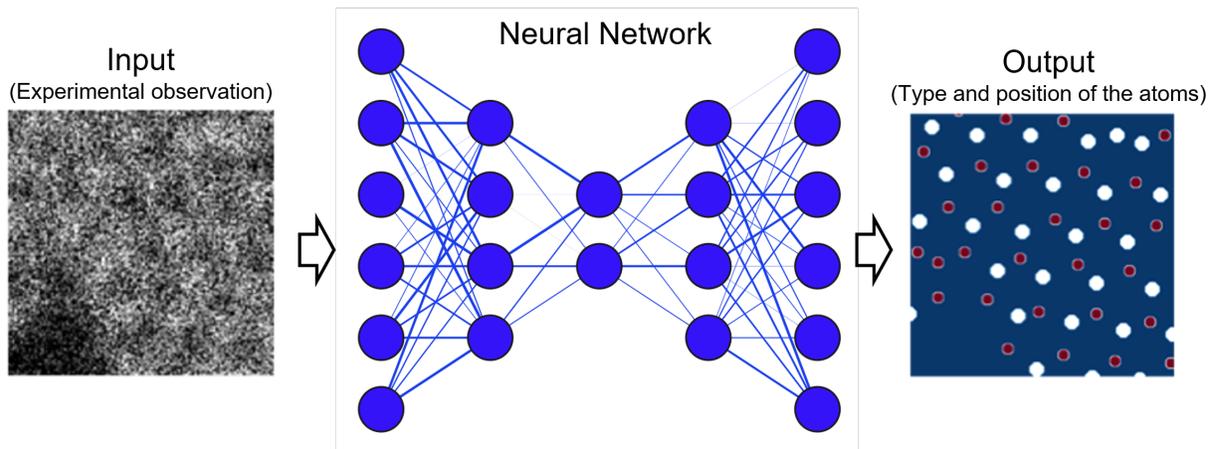

**Figure 1:** Deep neural network for analyzing atomically resolved data by performing a semantic image segmentation[7] on the atomic level. The network accomplishes two goals: i) it removes noise and ii) it separates different atomic species into different classes on the level of individual image pixels. The input image is the scanning transmission electron microscopy image of disordered atomic lattice of 2D boron nitride; the output is the atomic coordinates.

**When does ML work?**

From the early days of machine learning, it was repeatedly noted that ultimately ML and DL serve as universal interpolators, finding correlations between large datasets in multidimensional spaces. At the same time, the physical sciences are based on the notion of hypothesis-driven science, often using observations from a set of experiments to reveal correlations, explore causal relationships, and ultimately unveil the underpinning physical laws. Thus, when and how can ML and AI methods be used to explore physics?

We note that the pitfalls of the conventional correlative modelling and their consequences are well explored.[8,9] The classical examples include Simpson paradox[10], where for example it is possible that statistically, a certain drug can be beneficial for humans in general, but detrimental for both males and females. While in areas such as sociology, medicine, and economics the approaches to deal with these issues are well developed, this is generally not the case in the physical sciences. Notably, the use of complex machine learning models will not compensate for the incorrect causative attribution and would rather make the problem less obvious and more difficult to identify.

Here, we argue that the causal framework developed by Judea Pearl and expanded by Scholkopf, Mooij, and others provides a clear answer for these questions.[8,11,12] Generally, ML methods provide a universal and extremely powerful framework for analysis of physical problems when the causal chain is clearly known. The use of neural networks for the analysis of atomically resolved images[13] is causally-determined, since the point-like objects observed in electron microscope at this level of resolution can only be atoms. In comparison, these models will not generalize for all images, since large collection of sphere-like objects can also describe chain mail, cloth, meshes, structure of certain minerals, etc. Similarly, the use of generative adversarial networks for the analysis of the simulated 4D scanning transmission electron microscopy data (STEM), or classical back-propagation networks to identify Ising model parameters based on hysteresis loops is causally determined, since there is clear causal relationship between the inputs and outputs. At the same time, such trained networks can fail when applied to experimental data, since the instrument parameters are a confounding factor. In some cases, these can be accounted for by scaling and normalization, but not so in others, where calibration factors are numerous and the effect on the image is much more complex, and hence need to be calibrated in advance. Parenthetically, the outstanding success of DL learning methods as applied in the theoretical domain owes to the fact that the causal links there are explicit.

At the same time, ML methods can be expected to fail, and often fail, in cases where the causal links are uncertain. This includes multiple variants, including the presence of confounding factors that affect both (input) X and (output) Y, observational biases, etc. Correspondingly in these cases the ML model, no matter how good, will fail to predict and generalize since there are control factors outside of the model. For instance, if a material property is predicted by ML models on the basis of only local structure and global chemistry (and not local chemical environments), this can easily lead to erroneous predictions in cases where it is the local chemical environment driving the changes in the first place. Then, the question is, does machine learning here becomes useless? Interestingly, the answer is that is still extremely useful – as long as the model is used in the parameter space in which the confounding factors are constant and observations are made with the same biases.

So, what are the other areas for ML in physics, beyond the conditioned correlative models valid when the causal links are known or defined? One class of the models is those that explore the complexity of the dataset, either via manifold learning in purely data spaces, or symbolic reconstructions, or extraction of generative models. These models exploit the fact that physical laws are generally parsimonious. As an example, consider the use of neural networks with constraints placed on learned representations to answer a scientific hypothesis–that of a heliocentric solar system.[14] As analyzed by Lin, Tegmark and Rolnick,[15] the success of deep learning is inherently linked to the fact that most complex systems, including those in physics, are hierarchical and are drawn from a very small subset of all possible data distributions.

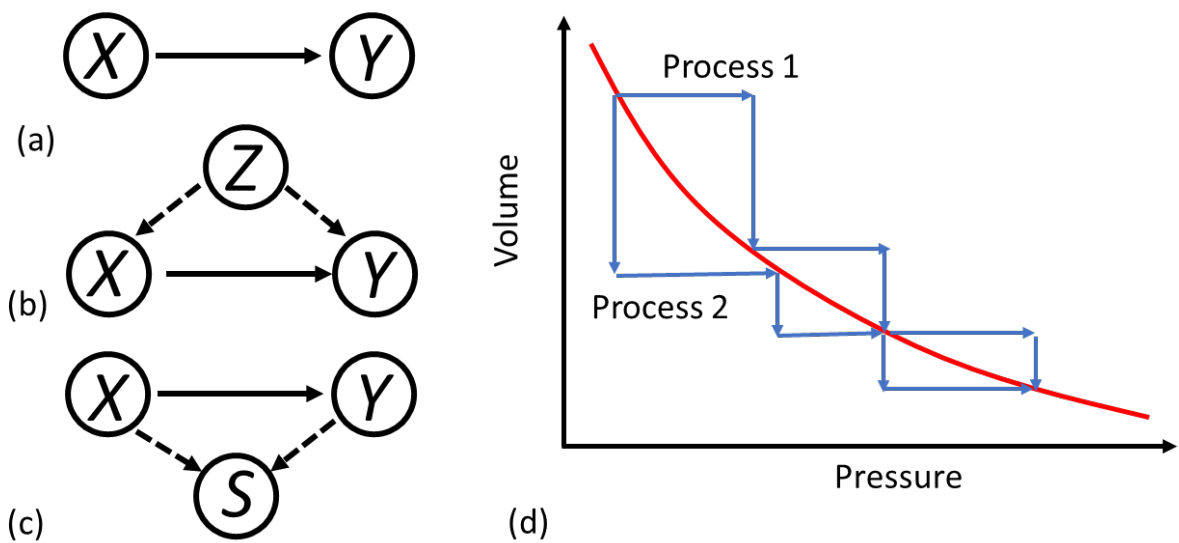

**Figure 2.** (a) Correlation can be used to analyze causative mechanisms only if there is a well-defined causal link between the variables. In the presence of (b) confounders or (c) observational bias analysis of correlations can result in fundamentally incorrect conclusions. For example, the correlation of the level of chocolate consumption and Nobel prize winning does not imply that chocolate can be used to increase scientific visibility; rather the same factors that enable higher consumption also lead to a higher probability of winning.[16] (d) Example of exploring causative mechanisms in physics. Observations will generate the correlation between pressure and volume. The analysis of the functional relationship between the two will yield the ideal gas law. With that, note that the knowledge of functional relationship is insufficient to analyze the causal mechanisms: does the pressure change cause changes in volume, or vice-versa?

*Learning Meaningful Representations: Looking for simplicity*

DL methods learn a representation of the inputs that is advantageous to the task that is required to be performed, which are sometimes referred to as 'features.' Are features learned by such networks physically reasonable or at all meaningful for materials scientists? After all, the predictions of a DNN may be highly accurate, but might have little to no extrapolation ability. This is because the features learned are the basis used for predictions of the model, and physically non-meaningful features can lead to highly inaccurate predictions for new, unseen data.

We argue that one method to aid the learning of *better* representations of systems is to incorporate principles from statistical physics. To be truly predictive, and not just interpolative, DNNs need to carry an internal representation of the physical system, which is ultimately given by its microstate probability distribution or partition function. Measured properties are then derived as specific projections *(i.e.,* coarse-graining) of the distribution. For example, for a 2D

Ising model the property of note (magnetization) can be derived in this way. The means to implement this can vary; however, the core aim is to ensure that the predictions are physically meaningful in terms of microstate probabilities. Moreover, one can employ regularization that is relevant for physics, in that we ensure such physical representations are, and must be expressed as only a small number of independent latent variables.[14] We note as an aside here that the links between neural networks and statistical physics, and the field of statistics more generally, go back at least three decades.[17]

*Adding context: Bayesian methods and prior knowledge*

Another major class of models are the Bayesian models. While DL requires large volumes of data and attempts to learn representations without the need for priors (beyond those encoded within the architecture design, such as convolutions which introduce spatial invariance), this is not the case for most physical problems. Indeed, the question of most importance is how best to incorporate prior knowledge of scientists within a data-driven approach.

The natural approach for incorporation of the past knowledge in the analysis is based on Bayesian methods, derived from the celebrated Bayes formula:

$$p(\theta_i|D) = \frac{p(D|\theta_i)p(\theta_i)}{p(D)} \qquad (1)$$

Here $D$ represents the new data, $p(D|\theta_i)$ is the likelihood that the data can be generated by the theory, i.e. the model, $i$, with parameters, $\theta$. The prior knowledge is represented by $p(\theta_i)$. Finally, $p(D)$ is the denominator that defined the total space of possible outcomes.

Despite the elegance and transparency of Bayesian approach, its adoption by many scientific communities has been rather slow. First, evaluation of denominator in Eq. 1 requires very high dimensional integrals and become feasible for experimentally relevant distributions only over the last decade. Secondly, the choice of the priors represents an obvious issue. Interestingly, in the physics field, domain knowledge is typically abundant, necessitating translating of past domain knowledge into the language of probability distribution functions. In a sense, Bayes formula represents the synergy of experimental science as a source of new data, domain expertise as source of priors, theory as a source of likelihoods, and high-performance computing necessary to address the associated computational challenges.

The adoption of Bayesian approaches allows us to systematically explore complex problems, fusing prior information from other sources. For example, in a scientific image processing task, can the neural network performance be improved based on knowledge on which functional groups are possible for specific materials class, and their relevant energies/probabilities? The combination of a convolutional neural networks with graphical models[18] (e.g. Markov random field) may allow incorporating prior knowledge about physiochemical properties of a system, such as a probability of realization of certain lattice-impurity configurations, into the decoding of experimental observations (see Figure 3). Deep learning models such as graph convolutional neural networks now also allow predicting materials properties directly from crystal lattice graphs.[19] However, this approach is currently limited to ideal periodic systems. Predicting the local property maps (e.g. distribution of local density of states) directly from the experimental observations (see Figure 3) is a major challenge.

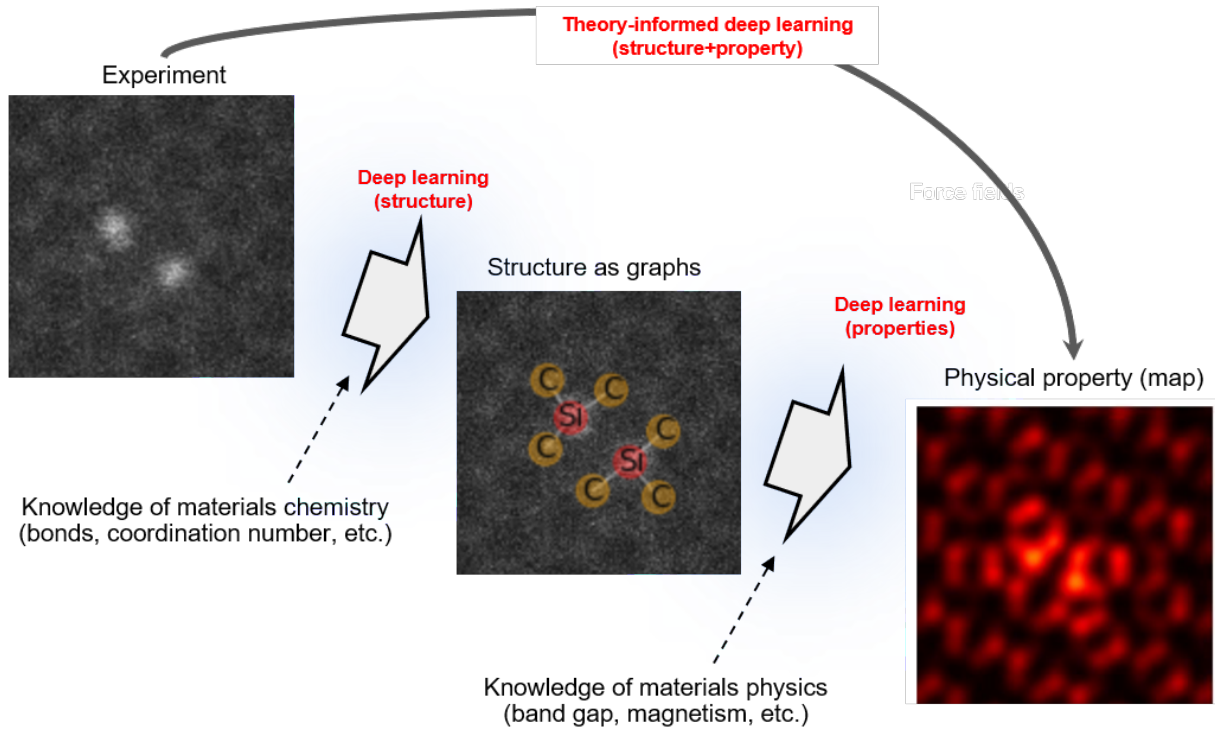

**Figure 3:** A simplified schematic showing analysis of scientific image data using a combination of deep learning and graph modelling for predicting materials (local) structure and properties. Knowledge can be injected at both the structural learning step, as well as the translation from structure to physical properties.

Other methods to leverage or couple physical models and machine learning have been proposed. For instance, the 'theory-driven data science' paradigm espoused by Karpatne et al.[20] describe several such approaches, including pre-training of networks with simulated data from physical models (so that when trained on real-world data, the networks are more likely to yield physically plausible results), and constrained optimization, where solutions must obey constraints such as being valid solutions to a partial differential equation. Determining the most effective methods to encode these relationships within deep neural networks remains an ongoing challenge, and invariably a tension between the flexibility of the model and the ability to learn physically meaningful relationships that underpin extrapolation ability will exist.

Of course, in some instances it may be better to avoid prior information entirely: for example, the AlphaZero[21] program mastered the games of Chess, Shogi and Go starting from random play and given no domain knowledge other than the game rules, and yet achieved super-human performance in all three. This can allow for new, novel strategies to emerge that humans may not have envisioned.[22] We foresee the utility of these approaches in particular to areas such as controlled materials synthesis, drug discovery[23], and other design spaces.[24]

*Data and DL Future*

Finally, we explore the changes in scientific community and infrastructure needed to make this deep learning transformation possible. Most of the critical algorithmic developments for deep learning, such as convolutional networks and back propagation, occurred decades ago.[25] Rather, it was the availability of large, labeled databases, and the ability to compute these huge volumes to enable network training, that were key factors in the current deep learning revolution.[26] As such, the development of open source libraries of materials data is an instrumental part, and a slew of recent reviews[27-29] touch on the need and benefits of these databases.

Similarly, adoption of machine learning tools, including basic knowledge and relevant programming skills by the broad scientific community becomes a must. A related issue is the availability and distribution of tested, well-documented codes. While GitHub and Jupyter notebooks[30] offer an effective means for code sharing, development, and universal access, the incentive system in fundamental science is heavily tilted towards publication as a primary measure of performance. Correspondingly, increasing visibility of code development and re-use, and ideally integrating codes into scientific publications becomes pertinent. Ultimately, data, code, and workflow sharing will become the primary pathway for collaboration and scientific knowledge dissemination, complementing and potentially surpassing archival publications.

Overall, the initial forays in machine learning across physical science communities have demonstrated the power of these methods in a variety of domains. But practical implementation will require additional work on adjusting the tools to match the problems presented in those areas. In our opinion, the integration of human domain expertise and causal inference with deep learning will be the crucial link to correctly harnessing and exploiting the benefits that DL and ML can provide. Most importantly, the merger of machine learning with classical hypothesis driven science can bring ML beyond the current correlative paradigms into larger fields of Bayesian and causal learning and establish connections to the materials world via automated experiment and open instrumental facilities, thus giving rise to fundamentally new ways of scientific research.

**Acknowledgements**

The work was supported by the U.S. Department of Energy, Office of Science, Materials Sciences and Engineering Division (S.V.K., L.V., R. K. V.). Research was conducted at the Center for Nanophase Materials Sciences, which also provided support (M.Z.) and is a US DOE Office of Science User Facility.

**Author contributions**
RKV wrote the outline and the majority of the perspective. SVK Provided the initial concept and participated in development of the ideas and assisted in manuscript preparation, and developed the causality section. LV wrote sections regarding connections between statistical physics and machine learning. MZ the wrote sections on contextual information and theory guidance.

**Competing Interests**
The authors declare that there are no competing interests

**Data availability**
No datasets were generated or analyzed during the current study